\documentclass[preprint]{JHEP3} 

\usepackage{epsfig,multicol}

\newcommand\fverb{\setbox\pippobox=\hbox\bgroup\verb}
\newcommand\fverbdo{\egroup\medskip\noindent%
			\fbox{\unhbox\pippobox}\ }
\newcommand\fverbit{\egroup\item[\fbox{\unhbox\pippobox}]}
\newbox\pippobox

\title{Anomalies on orbifolds with gauge symmetry breaking}

\author{Hyun Min Lee \\ 
	Physikalishes Institut, Universit\"at Bonn, 
	Nussallee 12, D-53115 Bonn, Germany\\
	E-mail: \email{minlee@th.physik.uni-bonn.de}
}

\preprint{\hepth{0211126}}	

\abstract{We embed two 4D chiral multiplets of opposite representations 
in the 5D $N=2$ $SU(N+K)$ gauge theory compactified on an orbifold
$S^1/(Z_2\times Z'_2)$. There are two types of orbifold boundary conditions 
in the extra dimension to obtain the 4D $N=1$ $SU(N)\times SU(K)\times U(1)$ 
gauge theory from the bulk: in Type I, one has the bulk gauge group 
at $y=0$ and the unbroken gauge group at $y=\pi R/2$ 
while in Type II, one has the unbroken gauge group at both fixed points.
In both types of orbifold boundary conditions, 
we consider the zero mode(s) as coming from a bulk $(K+N)$-plet 
and brane fields at the fixed point(s) with the unbroken gauge group. 
We check the consistency 
of this embedding of fields by the localized anomalies and the localized FI
terms. We show that the localized anomalies in Type I  
are cancelled exactly by the introduction of a bulk 
Chern-Simons term. On the other hand, in some class of Type II, 
the Chern-Simons term is 
not enough to cancel all localized anomalies even if they are globally 
vanishing. 
We also find that for the consistent embedding of brane fields, there appear 
only the localized log FI terms at the fixed point(s) with a $U(1)$ factor.  
}


\begin{document} 


\section{Introduction}

Recently the orbifold unification models in the existence of extra dimensions 
have 
drawn much attention due to their simplicity in performing the gauge symmetry 
breaking and the doublet-triplet splitting at the same time. 
The unwanted zero modes appearing in the unification models are projected out 
by boundary conditions in the extra dimension, i.e,  they get masses of order 
of the compactification scale. 
For instance,
the Minimal Supersymmetric Standard Model(MSSM) fields were obtained 
in the 5D SUSY SU(5) model where the extra dimension is compactified 
on a simple orbifold $S^1/(Z_2\times Z'_2)$\cite{kawamura,pointgroup,kkl0}. 
The idea was also taken in the model with the 5D $SU(3)$ electroweak 
unification with the TeV-sized extra dimension\cite{su3,kkl}, the possibility 
of which was first considered in the context of the string 
orbifolds\cite{anton}.

In the orbifold with gauge symmetry breaking, in general, 
in addition to the fixed point where the bulk gauge symmetry is operative, 
there exists a fixed point where only the unbroken gauge group is 
respected\cite{pointgroup}: for instance, 
$G_{SM}=SU(3)\times SU(2)\times U(1)$ in the case with the 5D SUSY SU(5) model
on $S^1/(Z_2\times Z'_2)$.  
Therefore, 
we can put a multiplet(so called a brane field) at that fixed point 
allowed by the representation of
the unbroken gauge group. 
In the more
realisic model constructions, there has been a lot of various possibilities
of having incomplete multiplets located at the orbifold fixed points
(or branes) with an unbroken gauge group\cite{pointgroup,kkl0,su3,kkl}.
For instance, in the 5D SU(5) GUT on $S^1/(Z_2\times Z'_2)$,
it has been shown that the $s-\mu$ puzzle can be understood
from the introduction of a split multiplet for $\bf 10$ of the second 
generation\cite{kkl0}
while the top-bottom mass hierarchy can be also explained with one Higgs in the
bulk and the other Higgs at the brane\cite{kkl0,kkl}. 
Moreover, introducing incomplete
multiplets for the quark setor is indispensible in the 5D SU(3) electroweak
unification on $S^1/(Z_2\times Z'_2)$\cite{su3,kkl}.

However, with incomplete multiplets, there could appear the localized 
gauge anomalies on the
independent orbifold boundaries\cite{scrucca,pilo,barbieri,scrucca2,kkl,nilles}.
It has been shown\cite{ah} that the abelian anomalies coming from a bulk 
field in 5D
are equally distributed at the fixed points by the half of its 4D anomaly
while the bulk Chern-Simons term\cite{ch} plays a role
in conveying localized anomaly at one fixed point 
to the other fixed point.
So, the 4D anomaly cancellation for zero modes is sufficient
for consistency. The anomaly analysis can be generalized to the case 
in the non-abelian gauge anomalies on the orbifold with gauge symmetry breaking.
It has been shown in 5D $SU(5)$ and $SU(3)$ orbifold unification 
models\cite{kkl}
that for fermion zero modes of the 4D anomaly-free combination, 
the integrated anomalies are absent but the localized anomalies
can be exactly cancelled by the introduction of a bulk Chern-Simons 
counter term.

In the 4D supersymmetric theory, it is known that 
the Fayet-Iliopoulos term(FI) also 
can be radiatively generated only for anomalous $U(1)$ gauge 
theories. This FI term could break supersymmetry and/or anomalous $U(1)$. 
However, the situation is somewhat different in orbifold models 
with a $U(1)$ factor.
In 5D $S^1/Z_2$ or $S^1/(Z_2\times Z'_2)$ orbifold 
with a $U(1)$ factor\cite{nilles2,barbieri,nilles,pomarol}, 
it has been recently shown that there is a possibility that 
the integrated FI term vanishes while there exist 
non-zero localized FI terms at the fixed points. 
These localized FI terms, however, do not affect either 4D supersymmetry or 
gauge symmetry since they can be absorbed by the non-vanishing vacuum 
expectation value of a real scalar field belonging 
to the bulk vector multiplet\cite{peskin,ah2,barbieri,nilles,pomarol}.
The net effect of the localized FI terms is the dynamical localization 
of the bulk zero mode and to make the bulk massive modes decoupled. 
Even in the case
where quadratically divergent FI terms are cancelled locally
by introducing at each fixed point a brane field with the half charge 
compared to that of a bulk field, 
the {\it logarithmically divergent} FI terms are equally distributed 
at both fixed points,
which also gives rise to the localization of the zero mode at both fixed 
points\cite{nilles}.

In this paper, we consider the gauge symmetry breaking due to the boundary 
conditions in the 5D $N=2$ SUSY $G=SU(N+K)$ gauge theory 
on $S^1/(Z_2\times Z'_2)$. 
In Type I, we have the full
gauge symmetry $G$ at $y=0$ and the unbroken gauge symmetry 
$H=SU(N)\times SU(K)\times U(1)$ at $y=\pi R/2$. On the other hand, in Type II,
we have the unbroken gauge group at both fixed points.
When we consider the zero modes coming from a bulk $(N+K)$-plet in both types
of orbifold boundary conditions, 
we show that the bulk fermion gives rise to the localized gauge
anomalies: $+\frac{1}{2}$ of $G^3$ gauge anomalies at $y=0$  
and $+\frac{1}{2}$ of $H^3$ gauge anomalies at $y=\pi R/2$ in Type I 
while $H^3$ gauge anomalies are equally split at both fixed points in Type II.  
Using this result, we find that 
addition of one $\bar K$-plet located at $y=\pi R/2$ in Type I leads to  
$-\frac{1}{2}$ of $H^3$ gauge anomalies at $y=\pi R/2$ 
while there remains $+\frac{1}{2}$ of $G^3$ gauge anomalies at $y=0$.
However, we also show that all the localized gauge anomalies are cancelled 
exactly by introducing a Chern-Simons(CS) 5-form with a jumping coefficient 
in the 5D action\cite{ch,ah,pilo}. On the other hand, in Type II with brane
fields at the same fixed point,
we show that a bulk CS term is not sufficient 
for cancelling the localized gauge anomalies. 

Secondly, we consider the localized FI terms in our model.
In Type I, the localized FI terms could appear only at $y=\pi R/2$ 
with a $U(1)$ factor. 
Since we have the $N=1$ supersymmetry without integrated gauge/gravitational 
anomalies at the zero mode level, the integrated FI term should vanish 
for consistency. Therefore,
it can be argued that the FI term should be absent locally in our 
model\cite{kkl}. On the other hand, in Type II, there could appear the FI terms
at both fixed points, which seems to be not necessarily zero with the condition
for no integrated FI term. However, since there is no gravitational counterpart
of CS term in GUT orbifolds, there should not be localized gravitational 
anomalies, which allows only a specific assigning of brane fields.
In this paper, we show that there exists 
only a non-vanishing log divergent FI term at $y=\pi R/2$ in Type I 
and log divergent FI terms at both fixed points in a consistent class 
of Type II, which still makes the integrated FI term to be zero. 
The log divergent FI term can be absorbed by a singular vacuum expectation
value(VEV) of the $U(1)$ gauge component of the real adjoint scalar field 
in the bulk without changing the mass spectrum\cite{nilles}. 

Our paper is organized as follows. In the next section, we give an introduction
to the gauge symmetry breaking on orbifolds by adopting  
the 5D SUSY $SU(N+K)$ gauge theory on $S^1/(Z_2\times Z'_2)$.
Then, in the section 3, for this GUT orbifold, 
we derive the detailed expression for the localized non-abelian anomalies
coming from a bulk fermion in the fundamental representation of $SU(N+K)$. 
The section 4 is devoted to the localization problem of a bulk fermion and the
cancellation of the localized gauge anomalies. 
In the section 5, we work out with 
the localized Fayet-Iliopoulos terms in our model. 
Then, we conclude the paper in the last section.

\section{Orbifold breaking of gauge symmetry}

Let us consider the five-dimensional SUSY $G=SU(N+K)$ gauge theory 
compactified
on an $S^1/(Z_2\times Z'_2)$ orbifold.
The fifth dimensional coordinate $y$ is compactified to a circle
$2\pi R\equiv 0$. Furthermore, the point $y=-a$ is identified to
$y=a$ ($Z_2$ symmetry) and the point $y=(\pi R/2)+a$ is identified
to $y=(\pi R/2)-a$ ($Z_2'$ symmetry).
Then, the fundamental region of
the extra dimension becomes the interval $[0,\frac{\pi R}{2}]$
between two fixed points $y=0$ and $y=\frac{\pi R}{2}$.

For the two $Z_2$ symmetries, one can define their actions $P$ and $P'$
within the configuration space of any bulk field:
\begin{eqnarray}
\phi(x,y)&\rightarrow& \phi(x,-y)=P\phi(x,y), \\
\phi(x,y')&\rightarrow& \phi(x,-y')=P'\phi(x,y')
\end{eqnarray}
where $y'\equiv y+\pi R/2$.
The $(P,P')$ actions can involve all the symmetries
of the bulk theory, for instance, the gauge symmetry and the R-symmetry
in the supersymmetric case.
In general, then, any bulk field $\phi$
can take one of four different Fourier expansions depending on their
pair of two $Z_2$ parities, $(i,j)$ as
\begin{equation}
\phi_{++}=\sum^\infty_{n=0}\sqrt{\frac{1}{2^{\delta_{n,0}}\pi R}}
\phi^{(2n)}_{++}(x^\mu)\cos\frac{2ny}{R}\label{mode1}
\end{equation}
\begin{equation}
\phi_{+-}=\sum^\infty_{n=0}\sqrt{\frac{1}{\pi R}}
\phi^{(2n+1)}_{+-}(x^\mu)\cos\frac{(2n+1)y}{R}\label{mode2}
\end{equation}
\begin{equation}
 \phi_{-+}=\sum^\infty_{n=0}\sqrt{\frac{1}{\pi R}}
\phi^{(2n+1)}_{-+}(x^\mu)\sin\frac{(2n+1)y}{R}\label{mode3}
\end{equation}
\begin{equation}
\phi_{--}=\sum^\infty_{n=0}\sqrt{\frac{1}{\pi R}}
\phi^{(2n+2)}_{--}(x^\mu)\sin\frac{(2n+2)y}{R}\label{mode4}
\end{equation}
where $x^\mu$ is the 4D space-time coordinate.

The minimal supersymmetry in 5D
corresponds to N=2 supersymmetry(or 8 supercharges)
in the 4D N=1 language.
Thus, a 5D chiral multiplet corresponds to an N=2 hypermultiplet consisting
of two N=1 chiral multiplets with opposite charges. Two 4D Weyl spinors make
up one 5D spinor. On the other hand,
a 5D vector multiplet corresponds to an N=2 vector multiplet composed of one
N=1 vector multiplet($V=(A_\mu,\lambda_1, D)\equiv V^q T^q$)
and one N=1 chiral multiplet
($\Sigma=((\Phi+iA_5)/\sqrt{2}, \lambda_2,F_\Sigma)\equiv \Sigma^q T^q$), 
which transforms in the adjoint representation of the bulk gauge 
group\footnote{We note $D=X_3-\partial_5\Sigma$ and 
$F_\Sigma=(X_1+iX_2)/\sqrt{2}$ 
in terms of the $SU(2)_R$ triplet $\vec X$ in $N=2$ supersymmetry.}.
Upon compactification, we consider the case
where one $Z_2$ breaks N=2 supersymmetry to N=1
while the other $Z_2$ breaks the bulk $G=SU(N+K)$ gauge group
to its subgroup $H=SU(N)\times SU(K)\times U(1)$.

For instance, a bulk hypermultiplet in the fundamental of $SU(N+K)$, 
which is composed of two chiral multiplets with opposite charges,
$H=(h,\psi, F_H)\equiv (H_1,H_2)^T$
and $H^c=(h^c,\psi^c, F_{H^c})\equiv (H^c_1, H^c_2)$,
transforms under $Z_2$ and $Z'_2$ identifications as
\begin{eqnarray}
H(x,-y)&=&\eta P H(x,y), \ \ \ H^c(x,-y)=- \eta H^c(x,y)P^{-1} \\
H(x,-y')&=&\eta' P' H(x,y'), \ \ \ H^c(x,-y')=-\eta'H^c(x,y') P^{\prime -1}
\end{eqnarray}
where both $\eta$ and $\eta'$ can take $+1$ or $-1$, and 
$P^2=P^{\prime 2}=I_{N+K}$ where $I_{N+K}$ is the $(N+K)\times (N+K)$ 
identity matrix. On the other hand, the bulk gauge multiplet
is transformed under the two $Z_2$ transformations respectively as
\begin{eqnarray}
V(x,-y)&=&PV(x,y)P^{-1}, \\
\Sigma(x,-y)&=&-P\Sigma(x,y)P^{-1}, \\
V(x,-y')&=&P'V(x,y')P^{\prime -1}, \\
\Sigma(x,-y')&=&-P'\Sigma(x,y')P^{\prime -1}.
\end{eqnarray}

Now let us consider two types of the parity assignment $(P,P')$ as
\begin{eqnarray}
{\rm Type\ \ I}:&~~&P= I_{N+K}, \ \ \ P'={\rm diag}(I_{K},-I_{N}), \label{parity}\\
{\rm Type\ \ II}:&~~& P=P'={\rm diag}(I_{K},-I_{N}).\label{parity2}
\end{eqnarray}
For both types of parity matrices, the $G=SU(N+K)$
gauge symmetry is broken down to $H=SU(N)\times SU(K)\times U(1)$
because $P'$($P$ also for Type II) 
does not commute with all the gauge generators of
$SU(N+K)$: $P' T^a P^{\prime -1}=T^a$ and $P' T^{\hat a}P^{\prime
-1}=-T^{\hat a}$ where $q=(a,\hat a)$ denote unbroken and broken
generators, respectively. 
However, the fixed point gauge groups are different: for Type I, 
the $G$ bulk gauge symmetry at $y=0$ and 
the $H$ unbroken gauge symmetry at $y=\pi R/2$; 
for Type II, the $H$ unbroken gauge symmetry at both $y=0$ and 
$y=\pi R/2$. Therefore, in either case, it is possible to put some incomplete
multiplets transforming only under the unbroken gauge group at the fixed 
point(s). 

For Type I with $\eta=\eta'=1$,
the bulk hypermultiplet in the fundamental of $SU(N+K)$ is split as follows
\begin{eqnarray}
&H_1^{(2n)}:&[(++);(1,K,\frac{1}{K})],\ {\rm mass}=2n/R \\
&H_2^{(2n+1)}:&[(+-);(N,1,-\frac{1}{N})],
\ {\rm mass}=(2n+1)/R \\
&H_2^{c(2n+1)}:& [(-+);(\overline {N},1,\frac{1}{N})],
\ {\rm mass}=(2n+1)/R \\
&H_1^{c(2n+2)}:&
[(--);(1,\overline{K},-\frac{1}{K})],\ {\rm mass}=(2n+2)/R
\end{eqnarray}
where the brackets [ ] contain the quantum numbers of $Z_2\times
Z_2'\times SU(N)\times SU(K)\times U(1)$. Consequently,
upon compactification, there appears
a zero mode only from the $K$-plet among the bulk field components while
other fields get massive.
Therefore, the bulk vector multiplet of
$SU(N+K)$ is divided into the different KK modes 
\begin{eqnarray}
&V^{a(n)}:&[(++);(N^2-1,1)+(1,K^2-1)
+(1,1)],\ {\rm mass}=2n/R \label{t1g1}\\
&V^{{\hat a}(2n+1)}:&[(+-);(N,K)
+(\overline{N},\overline{K})],
\ {\rm mass}=(2n+1)/R \label{t1g2}\\
&\Sigma^{{\hat a}(2n+1)}:&[(-+);(N,K)
+(\overline{N},\overline{K})],
\ {\rm mass}=(2n+1)/R \label{t1g3}\\
&\Sigma^{a(2n+2)}:&[(--);(N^2-1,1)+(1,K^2-1)
+(1,1)],
\ {\rm mass}=(2n+2)/R\label{t1g4}
\end{eqnarray}
where the brackets [ ] contain the quantum numbers of $Z_2\times
Z_2'\times SU(N)\times SU(K)$. Therefore, the orbifolding retains
only the $SU(N)\times SU(K)\times U(1)$ gauge multiplets as
massless modes $V^{a(0)}$ while the KK massive modes for unbroken
and broken gauge bosons are paired up separately. 

For Type II with $\eta=\eta'=1$, likewise, 
the bulk hypermultiplet is split as follows
\begin{eqnarray}
&H_1^{(2n)}:&[(++);(1,K,\frac{1}{K})],\ {\rm mass}=2n/R \\
&H_2^{(2n+2)}:&[(--);(N,1,-\frac{1}{N})],
\ {\rm mass}=(2n+2)/R \\
&H_2^{c(2n)}:& [(++);(\overline {N},1,\frac{1}{N})],
\ {\rm mass}=2n/R \\
&H_1^{c(2n+2)}:&
[(--);(1,\overline{K},-\frac{1}{K})].\ {\rm mass}=(2n+2)/R
\end{eqnarray}
Therefore, in this case, there appears a zero mode of $\bar N$-plet 
as well as a zero mode of $K$-plet. 
On the other hand, the bulk vector multiplet of $SU(N+K)$ is divided into
the different KK modes 
\begin{eqnarray}
&V^{a(n)}:&[(++);(N^2-1,1)+(1,K^2-1)
+(1,1)],\ {\rm mass}=2n/R \label{t2g1}\\
&V^{{\hat a}(2n+2)}:&[(--);(N,K)
+(\overline{N},\overline{K})],
\ {\rm mass}=(2n+2)/R \label{t2g2}\\
&\Sigma^{{\hat a}(2n)}:&[(++);(N,K)
+(\overline{N},\overline{K})],
\ {\rm mass}=2n/R \label{t2g3}\\
&\Sigma^{a(2n+2)}:&[(--);(N^2-1,1)+(1,K^2-1)
+(1,1)],
\ {\rm mass}=(2n+2)/R\label{t2g4}.
\end{eqnarray}
Thus, on top of the zero mode of the $H$ gauge multiplet, 
there exists a zero mode of the chiral multiplet $\Sigma^{\hat a}$ 
containing the broken gauge component of $A_5$. This new zero mode gets 
a radiative mass of order of the compactification scale and it can be 
identified with a Higgs multiplet, for instance, in the bulk $SU(3)$ gauge 
theory. 

Since each component of a gauge parameter
$\omega=\omega^q T^q$ has the same $Z_2$
parities as those of the corresponding gauge field,
the bulk gauge transformation is restricted as follows
\begin{eqnarray}
\delta A^a_M&=&\partial_M \omega^a+if^{abc}A^b_M \omega^c
+if^{a{\hat b}{\hat c}} A^{\hat b}_M\omega^{\hat c}, \label{gtransf1}\\
\delta A^{\hat a}_M &=&\partial_M \omega^{\hat a}+if^{{\hat a}{\hat b}c}
A^{\hat b}_M\omega^{c}+if^{{\hat a}b{\hat c}}A^b_M\omega^{\hat c}
\label{gtransf2}
\end{eqnarray}
where $f^{ab{\hat c}}$ and $f^{{\hat a}{\hat b}{\hat c}}$ are put to zero
for the parity invariance.
Particularly, since $\omega^{\hat a}$ takes the same parities $(+,-)((-,-))$ 
as $A^{\hat a}_\mu$ in Type I(II), 
the gauge transformation at $y=\pi R/2$(at both $y=0$ and $y=\pi R$) 
becomes the one of the unbroken gauge group $H$ from eq.~(\ref{gtransf1}).

\section{Non-abelian anomalies on orbifolds}

A 5D fermion is not chiral in the 4D language. However, after orbifold
compactification of the extra dimension, a chiral fermion can be
obtained as the zero mode of a bulk non-chiral fermion.
Then, the chiral fermion gives rise to the 4D gauge anomaly after
integrating out the extra dimension. For the case with the 5D $U(1)$ gauge
theory on $S^1/Z_2$\cite{ah} or $S^1/(Z_2\times Z'_2)$\cite{scrucca},
it was shown that the 4D gauge anomaly coming from a zero mode is equally
distributed at the fixed points.
In this section, we do the anomaly analysis in the case
with the 5D $SU(N+K)$ gauge theory compactified on our gauge symmetry
breaking orbifold, $S^1/(Z_2\times Z'_2)$.

Let us consider a four-component bulk fermion in the fundamental representation
of $SU(N+K)$. Then, the action is
\begin{eqnarray}
S=\int d^4 x \int_0^{2\pi R} dy\, \bar{\psi}(iD\hspace{-2.5mm}/
-\gamma_5 D_5-m(y))\psi \label{5daction}
\end{eqnarray}
where $D\hspace{-2.5mm}/=\gamma^\mu D_\mu $ and
$D_M=\partial_M+iA_M$. Here $m(y)$ is a mass term for the bulk fermion and 
$A_M=A^q_M T^q$ is a classical
 non-abelian gauge field.

With the assignments of $Z_2$ and $Z'_2$ parities to a $(N+K)$-plet
hypermultiplet in the previous section, the fermion field transforms as
\begin{eqnarray}
\psi(y)=\gamma_5 P\psi(-y), \ \ \ \psi(y')=\gamma_5 P'\psi(-y')
\end{eqnarray}
where $P$ and $P'$ are given by Eqs.~(\ref{parity}) and (\ref{parity2}), 
acting in the group space.
Invariance of the action under two $Z_2$'s gives rise to
the conditions for the mass function
\begin{eqnarray}
m(y)=-m(-y), \ \ \ m(y')=-m(-y').
\end{eqnarray}
And the gauge fields also transform under $Z_2$ as
\begin{eqnarray}
A_\mu(y)&=&P A_\mu(-y) P^{-1}, \ \ \ A_5(y)=-P A_5(-y)P^{-1},
\end{eqnarray}
and we replace ($y\rightarrow y'$, $P\rightarrow P'$) for $Z'_2$ action.

Then, with $\psi=\psi^1+\psi^2$, where $1$ and $2$ denotes
$K$-plet and $N$-plet components respectively, the fermion field
is decomposed into four independent chiral components
\begin{eqnarray}
\psi^1=\psi^1_L+\psi^1_R, \ \ \psi^2=\psi^2_L+\psi^2_R
\end{eqnarray}
where
\begin{eqnarray}
\gamma_5\psi^1_{L(R)}=\pm\psi^1_{L(R)}, \ \
\gamma_5\psi^2_{L(R)}=\pm\psi^2_{L(R)}.
\end{eqnarray}

First let us consider the case with Type I parity assignments.
Due to the parity assignments, $(\pm,\pm)$ for
$\psi^1_{L(R)}$ and $(\pm,\mp)$ for $\psi^2_{L(R)}$, we can expand
each Weyl fermion in terms of KK modes
\begin{eqnarray}
\psi^1_{L(R)}(x,y)&=&\sum_n \psi^1_{L(R)n}(x)\xi^{(\pm\pm)}_n(y), \\
\psi^2_{L(R)}(x,y)&=&\sum_n\psi^2_{L(R)n}(x)\xi^{(\pm\mp)}_n(y),
\end{eqnarray}
with 
\begin{eqnarray}
(-\partial_5+m(y))(\partial_5+m(y))\xi^{(+\pm)}_n(y)
&=&M^2_n\xi^{(+\pm)}_n(y),\\
(\partial_5+m(y))(-\partial_5+m(y))\xi^{(-\mp)}_n(y)
&=&M^2_n\xi^{(-\mp)}_n(y)
\end{eqnarray}
where $M_n$ is the $n$th KK mass.
Here we note that $\xi$'s make an orthonormal basis for the function
on $[0,2\pi R)$:
\begin{eqnarray}
\int_0^{2\pi R}dy\, \xi^{(\pm\pm)}_m(y)\xi^{(\pm\pm)}_n(y)
=\int_0^{2\pi R}dy\, \xi^{(\pm\mp)}_m(y)\xi^{(\pm\mp)}_n(y)=\delta_{mn},\\
\int_0^{2\pi R}dy\, \xi^{(++)}_m(y)\xi^{(\pm\mp)}_n(y)
=\int_0^{2\pi R}dy\, \xi^{(--)}_m(y)\xi^{(\pm\mp)}_n(y)=0.
\end{eqnarray}

Under the gauge $A_5=0$\footnote{The result will be not changed 
in the case without a gauge condition\cite{barbieri}}, 
inserting the mode sum of the fermion
into the 5D action, we obtain
\begin{eqnarray}
S&=&\int d^4 x\bigg[\sum_{n}\overline{\psi^1_n}
(i\partial\hspace{-2.5mm}/-M_{2n})
\psi^1_n+\sum_n\overline{\psi^2_n}(i\partial\hspace{-2.5mm}/-M_{2n-1})\psi^2_n
\nonumber \\
&-&\sum_{m,n}\bigg(V_{mn}(A^a)+V_{mn}(A^i)+V_{mn}(B)+V_{mn}(A^{\hat
a})\bigg)\bigg]
\end{eqnarray}
where $\psi^1_n=\psi^1_{Ln}+\psi^1_{Rn}$ for $n>0$
($\psi^1_0=\psi^1_{L0}$), $\psi^2_n=\psi^2_{Ln}+\psi^2_{Rn}$, and 
$V_{mn}$'s denote gauge vertex couplings. The $G=SU(N+K)$ gauge
fields($A=A^q T^q$) can be decomposed into 
\begin{eqnarray}
(N+K)^2-1\rightarrow (N^2-1,1)+(1,K^2-1)+(1,1)
+(N,K)+(\overline{N},\overline{K}),
\end{eqnarray}
that is, $A^aT^a(a=1,\cdots,
N^2-1)$, $A^iT^i(i=1,\cdots,K^2-1)$, 
$A^{(N+K)^2-1}T^{(N+K)^2-1}\equiv B T^B$ gauge fields 
for the $H=SU(N)\times SU(K)\times U(1)$ group, 
and $A^{\hat a }(t^{\hat a})_{\alpha r}\equiv X^{\alpha r}
(\alpha=1,\cdots,N;r=1,\cdots,K)$
gauge fields for the $G/H$ coset space, respectively. 
Here, broken group generators are related to $t^{\hat a}$ as 
\begin{eqnarray}
T^{\hat a}\equiv 
\left(\begin{array}{rr}
0 & t^{\hat a}\\ (t^{\hat a})^\dagger & 0
\end{array}\right).
\end{eqnarray}
Then, $V_{mn}$'s are given by the following:
\begin{eqnarray}
V_{mn}(A^a)&=&J^{\mu a }_{mn(+-)}{\cal A}^{a(+-)}_{mn\mu}
+J^{\mu a}_{mn(-+)}{\cal A}^{a(-+)}_{mn\mu} \nonumber \\
V_{mn}(A^i)&=&J^{\mu i}_{mn(++)}{\cal A}^{i(++)}_{mn\mu}
+J^{\mu i}_{mn(--)}{\cal A}^{a(--)}_{mn\mu} \nonumber \\
V_{mn}(B)&=&J^{\mu B}_{mn(++)}{\cal B}^{(++)}_{mn\mu}
+J^{\mu B}_{mn(--)}{\cal B}^{(--)}_{mn\mu}+J^{\mu B}_{mn(+-)}{\cal
B}^{(+-)}_{mn\mu} +J^{\mu B}_{mn(-+)}{\cal
B}^{(-+)}_{mn\mu}\nonumber\\
V_{mn}(A^{\hat a})&=&J^{\mu{\hat a}}_{mn(+)}{\cal A}^{{\hat a}(+)}_{mn\mu} 
+J^{\mu{\hat a}}_{mn(-)}{\cal A}^{{\hat a}(-)}_{mn\mu}
\end{eqnarray}
where
\begin{eqnarray}
{\cal A}^{a(\pm\mp)}_{mn\mu}&=&\int_0^{2\pi R}dy \,
\xi^{(\pm\mp)}_m(y)\xi^{(\pm\mp)}_n(y)A^{a}_\mu(x,y), \\
{\cal A}^{i(\pm\pm)}_{mn\mu}&=&\int_0^{2\pi R}dy \,
\xi^{(\pm\pm)}_m(y)\xi^{(\pm\pm)}_n(y)A^{i}_\mu(x,y), \\
{\cal B}^{(\pm\pm)}_{mn\mu}&=&\int_0^{2\pi R}dy \,
\xi^{(\pm\pm)}_m(y)\xi^{(\pm\pm)}_n(y)B_\mu(x,y), \\
{\cal B}^{(\pm\mp)}_{mn\mu}&=&\int_0^{2\pi R}dy \,
\xi^{(\pm\mp)}_m(y)\xi^{(\pm\mp)}_n(y)B_\mu(x,y), \\
{\cal A}^{{\hat a}(\pm)}_{mn\mu}&=&\int_0^{2\pi R}dy \,
\xi^{(\pm\mp)}_m(y)\xi^{(\pm\pm)}_n(y)A^{\hat a}_\mu(x,y)
\end{eqnarray}
and
\begin{eqnarray}
J^{\mu a}_{mn(\pm\mp)}=\overline{\psi^2_m}\gamma^\mu
P_{\pm}T^{a}\psi^2_n,& & J^{\mu
i}_{mn(\pm\pm)}=\overline{\psi^1_m}
\gamma^\mu P_{\pm}T^{i}\psi^1_n, \\
J^{\mu B}_{mn(\pm\pm)}=\overline{\psi^1_m}\gamma^\mu
P_{\pm}T^{B}_{K\times K}\psi^1_n,& &J^{\mu B}_{mn(\pm\mp)}=\overline{\psi^2_m}
\gamma^\mu P_{\pm}T^B_{N\times N}\psi^2_n,
\end{eqnarray}
\begin{eqnarray}
J^{\mu{\hat a}}_{mn(\pm)}=\overline{\psi^2_m}\gamma^\mu
P_{\pm}t^{\hat a}\psi^1_n+\overline{\psi^1_m}\gamma^\mu
P_{\pm}(t^{\hat a})^\dagger\psi^2_n
\end{eqnarray}
with $P_{\pm}=(1\pm\gamma_5)/2$. 
Here a decomposition of $T^B$ is understood 
such as $T^B={\rm diag.}(T^B_{N\times N},T^B_{K\times K})$. 
We note that the chiral
current for the $SU(N+M)$ gauge symmetry is split into chiral
currents coupled to the unbroken and broken gauge fields. 

Applying the classical equations of motion and the standard
results for the 4D chiral anomalies\cite{ah,pilo,barbieri}, 
we can derive the anomalies
for the chiral currents classified above. 
By making an inverse Fourier-transformation by the convolution of the bulk 
eigenmodes, the 5D gauge vector current 
$J^{Mq}={\overline\psi}\gamma^MT^q\psi$ is given by 
\begin{eqnarray}
J^{\mu a}(x,y)&=&\sum_{m,n}(\xi^{(+-)}_m\xi^{(+-)}_n J^{\mu a}_{mn(+-)}
+\xi^{(-+)}_m\xi^{(-+)}_n J^{\mu a}_{mn(-+)}),\\
J^{\mu i}(x,y)&=&\sum_{m,n}(\xi^{(++)}_m\xi^{(++)}_n J^{\mu i}_{mn(++)}
+\xi^{(--)}_m\xi^{(--)}_n J^{\mu i}_{mn(--)}), \\
J^{\mu B}(x,y)&=&\sum_{m,n}(\xi^{(++)}_m\xi^{(++)}_n J^{\mu B}_{mn(++)} 
+\xi^{(--)}_m\xi^{(--)}_n J^{\mu B}_{mn(--)} \nonumber \\
&+&\xi^{(+-)}_m\xi^{(+-)}_n J^{\mu B}_{mn(+-)}
+\xi^{(-+)}_m\xi^{(-+)}_n J^{\mu B}_{mn(-+)}),\\
J^{\mu{\hat a}}(x,y)&=&\sum_{m,n}(\xi^{(+-)}_m\xi^{(++)}_n 
J^{\mu{\hat a}}_{mn(+)}+\xi^{(-+)}_m\xi^{(--)}_n J^{\mu{\hat a}}_{mn(-)})
\end{eqnarray}
and we can construct $J^{5 q}$ similarly. 
Consequently, it turns out 
that the divergence of the 5D gauge vector current is given 
in terms of the 4D gauge anomalies as 
\begin{eqnarray}
(D_M J^{M})^a(x,y)&=&f_2(y)({\cal Q}^a(A) +{\cal Q}^a(X)), \\
(D_M J^{M})^i(x,y)&=&f_1(y)({\cal Q}^i(A) +{\cal Q}^i(X)), \\
(D_M J^{M})^B(x,y)&=&f_1(y)({\cal Q}^B_+(A)+{\cal Q}^B_+(X)) 
+f_2(y)({\cal Q}^B_-(A)+{\cal Q}^B_-(X)), \\
(D_M J^{M})^{\hat a}(x,y)&=&f_1(y)({\cal Q}^{\hat a}_1(X)
+{\cal Q}^{\hat a}_+(X))+f_2(y)({\cal Q}^{\hat a}_2(X)
+{\cal Q}^{\hat a}_-(X))
\end{eqnarray}
where
\begin{eqnarray}
f_1(y)&=&\sum_n\bigg[(\xi^{(++)}_n(y))^2 -(\xi^{(--)}_n(y))^2\bigg]
=\frac{1}{4}\sum_n \delta(y-\frac{n\pi R}{2}), \\
f_2(y)&=&\sum_n\bigg[(\xi^{(+-)}_n(y))^2 -(\xi^{(-+)}_n(y))^2\bigg]
=\frac{1}{4}\sum_n (-1)^n\delta(y-\frac{n\pi R}{2}).
\end{eqnarray}
The localized gauge anomalies ${\cal Q}$'s are composed of two large parts: 
anomalies for unbroken group components and broken group components of the 5D
vector current. The anomalies for unbroken group components involve 
not only unbroken gauge fields 
\begin{eqnarray}
{\cal Q}^a(A)&=&\frac{1}{32\pi^2}(D^{abc}F^b_{\mu\nu}{\tilde F}^{c\mu\nu}(x,y)
+D^{abB} F^b_{\mu\nu}{\tilde F}^{B\mu\nu}(x,y)), \\
{\cal Q}^i(A)&=&\frac{1}{32\pi^2}D^{ijB}F^j_{\mu\nu}
{\tilde F}^{B\mu\nu}(x,y)+\frac{1}{32\pi^2}D^{ijk}F^j_{\mu\nu}
{\tilde F}^{k\mu\nu}(x,y), \\
{\cal Q}^B_+(A)&=&\frac{1}{32\pi^2}{\rm Tr}(T^B_{K\times K})^3
F^B_{\mu\nu}{\tilde F}^{B\mu\nu}(x,y)\nonumber \\
&+&\frac{1}{64\pi^2}{\rm Tr}(\{T^B_{K\times K},T^i\}T^j)F^i_{\mu\nu}
{\tilde F}^{j\mu\nu}(x,y), \\ 
{\cal Q}^B_-(A)&=&\frac{1}{32\pi^2}{\rm Tr}(T^B_{N\times N})^3
F^B_{\mu\nu}{\tilde F}^{B\mu\nu}(x,y) \nonumber \\
&+&\frac{1}{64\pi^2}{\rm Tr}(\{T^B_{N\times N},T^a\}T^b)F^a_{\mu\nu}
{\tilde F}^{b\mu\nu}(x,y), \\
{\cal Q}^B_+(A)+{\cal Q}^B_-(A)&=&\frac{1}{32\pi^2}(D^{BBB}
F^B_{\mu\nu}{\tilde F}^{B\mu\nu}(x,y)
+D^{Bij}F^i_{\mu\nu}{\tilde F}^{j\mu\nu}(x,y) \nonumber \\
&+&D^{Bab}F^a_{\mu\nu}{\tilde F}^{b\mu\nu}(x,y))\equiv {\cal Q}^B(A), 
\end{eqnarray}
but also broken gauge fields 
\begin{eqnarray}
{\cal Q}^a(X)&=&\frac{1}{64\pi^2}{\rm Tr}(T^a t^{\hat b}(t^{\hat c})^\dagger)
F^{\hat b}_{\mu\nu}{\tilde F}^{{\hat c}\mu\nu}(x,y)
=\frac{1}{32\pi^2}D^{a{\hat b}{\hat c}}
F^{\hat b}_{\mu\nu}{\tilde F}^{{\hat c}\mu\nu}(x,y),\\
{\cal Q}^i(X)&=&\frac{1}{64\pi^2}{\rm Tr}(T^i (t^{\hat b})^\dagger t^{\hat c})
F^{\hat b}_{\mu\nu}{\tilde F}^{{\hat c}\mu\nu}(x,y)
=\frac{1}{32\pi^2}D^{i{\hat b}{\hat c}}
F^{\hat b}_{\mu\nu}{\tilde F}^{{\hat c}\mu\nu}(x,y), \\
{\cal Q}^B_+(X)&=&\frac{1}{64\pi^2}{\rm Tr}
(\{T^B,(t^{\hat b})^\dagger\}t^{\hat c})
F^{\hat b}_{\mu\nu}{\tilde F}^{{\hat c}\mu\nu}(x,y),  \\
{\cal Q}^B_-(X)&=&\frac{1}{64\pi^2}{\rm Tr}
(\{T^B,t^{\hat b}\}(t^{\hat c})^\dagger)
F^{\hat b}_{\mu\nu}{\tilde F}^{{\hat c}\mu\nu}(x,y), \\
{\cal Q}^B_+(X)+{\cal Q}^B_-(X)
&=&\frac{1}{32\pi^2}D^{B{\hat b}{\hat c}}
F^{\hat b}_{\mu\nu}{\tilde F}^{{\hat c}\mu\nu}(x,y)\equiv {\cal Q}^B(X).
\end{eqnarray}
On the other hand, the anomalies for broken group components of the 5D vector 
current become 
\begin{eqnarray}
{\cal Q}^{\hat a}_1(X)&=&\frac{1}{64\pi^2}{\rm Tr}
((\{t^{\hat a},(t^{\hat b })^\dagger\}
+\{(t^{\hat a})^\dagger,t^{\hat b}\})T^a)
F^{\hat b}_{\mu\nu}{\tilde F}^{a\mu\nu}(x,y) \nonumber \\
&=&\frac{1}{32\pi^2}D^{{\hat a}{\hat b}a}
F^{\hat b}_{\mu\nu}{\tilde F}^{a\mu\nu}(x,y),\\
{\cal Q}^{\hat a}_2(X)&=&\frac{1}{64\pi^2}{\rm Tr}
((\{t^{\hat a},(t^{\hat b})^\dagger\}
+\{(t^{\hat a})^\dagger,t^{\hat b}\})T^i)
F^{\hat b}_{\mu\nu}{\tilde F}^{i\mu\nu}(x,y)  \nonumber \\
&=&\frac{1}{32\pi^2}D^{{\hat a}{\hat b}i}
F^{\hat b}_{\mu\nu}{\tilde F}^{i\mu\nu}(x,y),\\
{\cal Q}^{\hat a}_+(X)&=&\frac{1}{64\pi^2}{\rm Tr}
(((t^{\hat a})^\dagger t^{\hat b}+(t^{\hat b })^\dagger t^{\hat a})
T^B_{K\times K})F^{\hat b}_{\mu\nu}{\tilde F}^{B\mu\nu}(x,y) \nonumber \\
&=&\frac{1}{64\pi^2}{\rm Tr} (\{T^{\hat a},T^{\hat b}\}T^B_{K\times K})
F^{\hat b}_{\mu\nu}{\tilde F}^{B\mu\nu}(x,y), \\
{\cal Q}^{\hat a}_-(X)&=&\frac{1}{64\pi^2}{\rm Tr}
((t^{\hat a} (t^{\hat b})^\dagger+t^{\hat b} (t^{\hat a})^\dagger)
T^B_{N\times N})F^{\hat b}_{\mu\nu}{\tilde F}^{B\mu\nu}(x,y) \nonumber\\ 
&=&\frac{1}{64\pi^2}{\rm Tr} (\{T^{\hat a},T^{\hat b}\}T^B_{N\times N})
F^{\hat b}_{\mu\nu}{\tilde F}^{B\mu\nu}(x,y), \\
{\cal Q}^{\hat a}_+(X)+{\cal Q}^{\hat a}_-(X)&=&
\frac{1}{32\pi^2}D^{{\hat a}{\hat b}B}F^{\hat b}_{\mu\nu}
{\tilde F}^{B\mu\nu}(x,y) \equiv {\cal Q}^{\hat a}_3(X)
\end{eqnarray}
In all the expressions for the anomalies above,  
we note that $D^{abc}$ denotes the symmetrized trace of group generators
\begin{eqnarray}
D^{abc}=\frac{1}{2}{\rm Tr}(\{T^a,T^b\}T^c)
\end{eqnarray} 
and other $D$ symbols with different group idices are similarly understood. 

As a result, we find that a bulk fermion gives rise to the localized gauge 
anomalies for all gauge components of the 5D vector current.  
Since the broken gauge fields vanish at $y=\pi R/2$ due to their
boundary conditions, the localized gauge anomalies 
at $y=\pi R/2$ are only ${\cal Q}(A)$'s, i.e., the $H^3$ gauge anomalies. 
However, at the other fixed point $y=0$, in addition to ${\cal Q}(A)$'s, 
there also appear the localized gauge 
anomalies ${\cal Q}(X)$'s, so we obtain the localized anomalies of $G^3$
at $y=0$.
Restricting to the region $[0,2\pi R)$, 
we can rewrite the divergence of the 5D vector current as
\begin{eqnarray}
(D_M J^{M})^a(x,y)&=&\frac{1}{2}\bigg(\delta(y)-\delta(y-\frac{\pi R}{2})\bigg)
{\cal Q}^a(A) 
+\frac{1}{2}\delta(y){\cal Q}^a(X), \label{am1}\\
(D_M J^{M})^i(x,y)&=&\frac{1}{2}\bigg(\delta(y)+\delta(y-\frac{\pi R}{2})\bigg)
{\cal Q}^i(A) 
+\frac{1}{2}\delta(y){\cal Q}^i(X), \label{am2}\\
(D_M J^{M})^B(x,y)&=&\frac{1}{2}\bigg(\delta(y)+\delta(y-\frac{\pi R}{2})\bigg)
{\cal Q}^B_+(A)
+\frac{1}{2}\bigg(\delta(y)-\delta(y-\frac{\pi R}{2})\bigg){\cal Q}^B_-(A)
\nonumber \\
&+&\frac{1}{2}\delta(y){\cal Q}^B(X),\label{am3}\\
(D_M J^{M})^{\hat a}(x,y)&=&\frac{1}{2}\delta(y) {\cal Q}^{\hat a}(X)\label{am4}
\end{eqnarray} 
where ${\cal Q}^{\hat a}(X)\equiv 
{\cal Q}^{\hat a}_1(X)+{\cal Q}^{\hat a}_2(X)+{\cal Q}^{\hat a}_3(X)$.

For the case with Type II parity assignments, 
making the mode expansions of the bulk fermion as
\begin{eqnarray}
\psi^1_{L(R)}(x,y)&=&\sum_n\psi^1_{L(R)n}(x)\xi^{(\pm\pm)}_n(y), \\
\psi^2_{L(R)}(x,y)&=&\sum_n\psi^2_{L(R)n}(x)\xi^{(\mp\mp)}_n(y)
\end{eqnarray}
which make up a Dirac fermion at each KK level as 
$\psi^1_n=\psi^1_{Ln}+\psi^1_{Rn}$ and $\psi^2_n=\psi^2_{Ln}+\psi^2_{Rn}$
for $n>0$($\psi^1_0=\psi^1_{L0}$ and $\psi^2_0=\psi^2_{R0}$),
and following the similar procedure as before,
a bulk fermion gives rise to the divergence of the 5D vector current as
\begin{eqnarray}
(D_M J^{M})^a(x,y)&=&-\frac{1}{2}\bigg(\delta(y)+\delta(y-\frac{\pi R}{2})\bigg)
{\cal Q}^a(A), \label{2am1}\\
(D_M J^{M})^i(x,y)&=&\frac{1}{2}\bigg(\delta(y)+\delta(y-\frac{\pi R}{2})\bigg)
{\cal Q}^i(A), \label{2am2}\\
(D_M J^{M})^B(x,y)&=&\frac{1}{2}\bigg(\delta(y)+\delta(y-\frac{\pi R}{2})\bigg)
({\cal Q}^B_+(A)-{\cal Q}^B_-(A)),\label{2am3}\\
(D_M J^{M})^{\hat a}(x,y)&=&0. \label{2am4}
\end{eqnarray}
Therefore, we find that there is no localized anomalies related to $X,Y$ gauge
bosons and the $H$ anomalies coming from the zero modes of 
$K$-plet and $\bar N$-plet are equally distributed on the orbifold fixed points.

\section{Localization of a bulk field and anomaly problem}

As shown in the section 2, we can freely put some brane fields  
consistently with the local gauge symmetries at the fixed points: 
a brane field at $y=0$ should be a representation of $SU(N+K)$ while
a brane field at $y=\pi R/2$ should be a representation 
of $SU(N)\times SU(K)\times U(1)$. Since we assume that a bulk fermion gives 
rise to a $K$-plet as the zero mode and we want to have 
the anomaly-free theory at least at the zero mode level, 
we can only put a brane field of $\bar K$-plet at $y=\pi R/2$. 
This introduction of an incomplete brane multiplet is
sufficient for the 4D anomaly-free theory at low energies but it could be 
inconsistent due to the existence of the localized gauge anomalies 
on the boundaries of the extra dimension. In this section, we consider
the localization of a bulk fermion with a kink mass and subsequently deal 
with the appearing anomaly problem by using the results in the previous 
section. 

It was shown in the literature that the localization of a bulk fermion 
can be realized by introducing a kink mass in the Lagrangian and 
even a brane fermion is possible in the limit of a kink mass being 
infinite\cite{ah}. 
In the 5D $U(1)$ gauge theory on $S^1/Z_2$ with a single bulk fermion, 
as a result of introducing an infinite kink mass,
the anomaly contribution from a bulk fermion on the boundaries of the extra
dimension was interpreted as the sum 
of contributions from a brane fermion and a parity-violating 
Chern-Simon term in 5D\cite{ah}. 
In other words, as a kink mass becomes infinite, 
heavy KK modes are decoupled but their effects remain as a local counterterm
such as the 5D Chern-Simon term. The similar observation has been made
for the non-abelian anomalies on orbifolds\cite{kkl}. 

In our case with gauge symmetry breaking on orbifolds, 
an infinite kink mass, depending on its sign, 
could give rise to the localization of the unwanted bulk modes 
as massless modes\cite{barbieri,pomarol,hebecker} in Type I. For instance, 
a positive(negative) infinite kink mass for the even modes 
($(+,+)$ and $(-,-)$) 
gives rise to a localization of the massless mode for $(+,+)$ 
at $y=0(y=\pi R/2)$. 
On the other hand, 
a positive infinite kink mass for the odd modes ($(+,-)$ and $(-,+)$) 
could lead to new massless modes localized at $y=0$ and $y=\pi R/2$, 
respectively. 
Suppose that there are the universal(preserving the bulk gauge symmetry) 
kink masses for even and odd modes, i.e., 
$m(y)=M\epsilon(y)I_{(N+K)\times (N+K)}$ in eq.~(\ref{5daction}) 
where $\epsilon(y)$ is the sign function with periodicity $\pi R$. 
Then, in order to avoid unwanted massless modes in Type I,
we only have to take the sign of $M$ to be negative. 
For instance, when we introduce a bulk multiplet $\overline{(N+K)}$ 
with $M\rightarrow -\infty$, 
we obtain a massless $\bar K$-plet only from the $(+,+)$ mode, 
which is localized at $y=\pi R/2$, while other modes get decoupled 
from the theory. 
Thus, in this respect, a brane $\bar K$-plet is naturally realized from a bulk 
complete multiplet in the field theoretic limit. 
In this process of localization, we find that 
the consistency with the incomplete brane field can be guaranteed 
with introducing a 5D Chern-Simons term\cite{kkl}, 
which would be interpreted as the 
effects from the decoupled heavy modes\cite{ch,ah,pilo}. 
On the other hand, in Type II, a bulk $\overline {(N+K)}$ with $M>0(M<0)$ 
also gives rise to the localization of two zero modes at different branes: 
the zero mode of $\bar K$-plet at $y=0$($y=\pi R/2$) and the zero mode 
of $N$-plet at $y=\pi R/2$($y=0$). However, in this case, it is not possible 
to localize the two brane fields at the same brane in a field theory. 
This is related to the inconsistency of a class of Type II 
in view of localized gauge and gravitational anomalies which are
not cancelled by the CS contributions as will be shown later.    

Let us consider the local anomaly cancellation in both Types I and II 
of GUT orbifolds. To begin with, in Type I,
we can introduce a brane $\bar K$-plet at $y=\pi R/2$ on top of the bulk 
$(N+K)$-plet.
Then, it gives rise to 4D
gauge anomalies such as $-{\cal Q}^i(A)$ and $-{\cal Q}^B_+(A)$ at that fixed 
point, which cancel
the 4D global anomalies coming from the zero mode of the bulk $K$-plet. 
Therefore, the resultant divergence of the 5D vector current is changed to
\begin{eqnarray}
(D_M J^{M})^a(x,y)&=&\frac{1}{2}\bigg(\delta(y)-\delta(y-\frac{\pi R}{2})\bigg)
{\cal Q}^a(A)+\frac{1}{2}\delta(y){\cal Q}^a(X), \label{bbanomaly1}\\
(D_M J^{M})^i(x,y)&=&\frac{1}{2}\bigg(\delta(y)-\delta(y-\frac{\pi R}{2})\bigg)
{\cal Q}^i(A)+\frac{1}{2}\delta(y){\cal Q}^i(X), \label{bbanomaly2}\\
(D_M J^{M})^B(x,y)&=&\frac{1}{2}\bigg(\delta(y)-\delta(y-\frac{\pi R}{2})\bigg)
{\cal Q}^B(A)
+\frac{1}{2}\delta(y){\cal Q}^B(X), \label{bbanomaly3}\\
(D_M J^{M})^{\hat a}(x,y)&=&\frac{1}{2}\delta(y)
{\cal Q}^{\hat a}(X) \label{bbanomaly4}.
\end{eqnarray}
Here we observe that the total localized gauge anomalies only involving
the unbroken gauge group(${\cal Q}(A)$'s) appear 
in the combination of $(\delta(y)-\delta(y-\pi R/2))$, so their integrated
gauge anomalies vanish.  
On the other hand, the anomalies involving 
broken gauge fields(${\cal Q}(X)$'s) 
remain nonzero even after integration because ${\cal Q}(X)$'s 
are nonzero only at $y=0$. This asymmetric localization of  
${\cal Q}(X)$'s reflects the difference between two fixed point groups 
in Type I. 

On the other hand, in Type II, we can introduce two brane incomplete 
fields($\bar K$-plet and $N$-plet) for no global anomalies
in different ways since both fixed points have only the unbroken gauge group
operative. Two brane fields could be located at the same brane or different 
branes. First let us consider the case that two brane fields 
are located at different branes, i.e. $\bar K$-plet at $y=\pi R/2$($y=0$) 
and $N$-plet at $y=0$($y=\pi R/2$). Then, the nonvanishing localized anomalies 
become
\begin{eqnarray}
(D_M J^{M})^a(x,y)&=&\pm\frac{1}{2}\bigg(\delta(y)-\delta(y-\frac{\pi R}{2})\bigg)
{\cal Q}^a(A), \label{type2a}\\
(D_M J^{M})^i(x,y)&=&\pm\frac{1}{2}\bigg(\delta(y)-\delta(y-\frac{\pi R}{2})\bigg)
{\cal Q}^i(A), \label{type2b}\\
(D_M J^{M})^B(x,y)&=&\pm\frac{1}{2}\bigg(\delta(y)-\delta(y-\frac{\pi R}{2})\bigg)
{\cal Q}^B(A).
\end{eqnarray}
Thus, the integrated anomalies vanish as expected but there exist localized
gauge anomalies involving the unbroken gauge group.
Secondly, in case that two brane fields are located at the same branes
$y=0$($y=\pi R$), we get the total localized anomalies as
\begin{eqnarray}
(D_M J^{M})^a(x,y)&=&\pm\frac{1}{2}\bigg(\delta(y)-\delta(y-\frac{\pi R}{2})\bigg)
{\cal Q}^a(A), \\
(D_M J^{M})^i(x,y)&=&\mp\frac{1}{2}\bigg(\delta(y)-\delta(y-\frac{\pi R}{2})\bigg)
{\cal Q}^i(A), \\
(D_M J^{M})^B(x,y)&=&\pm\frac{1}{2}\bigg(\delta(y)-\delta(y-\frac{\pi R}{2})\bigg)({\cal Q}^B_-(A)-{\cal Q}^B_+(A)).
\end{eqnarray} 
In this case, we have no integrated anomalies either but the structure 
of localized anomalies are different from the case with two brane fields 
at different branes. 

The existence of the localized gauge anomalies in either type of GUT orbifolds 
could make the theory with the unbroken gauge group anomalous.  
However, these localized gauge anomalies can be 
exactly cancelled with the introduction of a Chern-Simons(CS) 5-form 
$Q_5[A=A^q T^q]$ with a jumping coefficient in the action\cite{kkl} 
\begin{eqnarray}
{\cal L}_{CS}=-\frac{1}{96\pi^2}\epsilon(y)Q_5[A]
\end{eqnarray}  
where $\epsilon(y)$ is the sign function with periodicity $\pi R$ and 
\begin{eqnarray}
Q_5[A]={\rm Tr}\bigg(AdAdA+\frac{3}{2}A^3 dA+\frac{3}{5}A^5\bigg).
\end{eqnarray}
The parity-odd function $\epsilon(y)$ in front of $Q_5$ is necessary 
for the parity invariance because
$Q_5$ is a parity-odd quantity according to our parity assignments for bulk
gauge fields, eqs.~(\ref{t1g1})-(\ref{t1g4}) for Type I 
and eqs.~(\ref{t2g1})-(\ref{t2g4}) for Type II. 
Under the gauge transformation 
$\delta A=d\omega+[A,\omega]\equiv D\omega$, 
\begin{eqnarray}
\delta Q_5=Q^1_4[\delta A,A]={\rm str}
\bigg(D\omega \,d(AdA+\frac{1}{2}A^3)\bigg)
\end{eqnarray}
where str means the symmetrized trace and the restricted gauge transformation
in eqs.~(\ref{gtransf1}) and (\ref{gtransf2}) is understood. 
Then, due to the sign function in front of $Q_5$, the variation 
of the Chern-Simons action gives rise to the 4D consistent anomalies on the 
boundaries for Type I and II, respectively,
\begin{eqnarray}
\delta {\cal L}_{CS}&=&\frac{1}{48\pi^2}(\delta(y)-\delta(y-\frac{\pi R}{2}))
\,\epsilon^{\mu\nu\rho\sigma}\sum_{q=a,i,B}\omega^q{\rm str}(T^q\partial_\mu
(A_\nu\partial_\rho A_\sigma+\frac{1}{2}A_\nu A_\rho A_\sigma)) 
\nonumber \\
&+&\frac{1}{48\pi^2}\delta(y)\,\epsilon^{\mu\nu\rho\sigma}
\omega^{\hat a}{\rm str}
(T^{\hat a}\partial_\mu
(A_\nu\partial_\rho A_\sigma+\frac{1}{2}A_\nu A_\rho A_\sigma)), \\
\delta {\cal L}_{CS}&=&\frac{1}{48\pi^2}(\delta(y)-\delta(y-\frac{\pi R}{2}))
\,\epsilon^{\mu\nu\rho\sigma}\sum_{q=a,i,B}\omega^q{\rm str}(T^q\partial_\mu
(A_\nu\partial_\rho A_\sigma+\frac{1}{2}A_\nu A_\rho A_\sigma)).
\end{eqnarray} 
The consistent anomalies for Type I we obtained here 
can be changed to the covariant 
anomalies\cite{bardeen} by regarding the covariant non-abelian gauge current 
$J^q_\mu$ as being redefined from a non-covariant gauge current 
${\tilde J}^q_\mu$ as
\begin{eqnarray}
J^q_\mu(x,y)={\tilde J}^q_\mu(x,y)+U^q_\mu(x,y)
\end{eqnarray}
where 
\begin{eqnarray}
U^{q=(a,i,B)}_\mu&=&-\frac{1}{96\pi^2}(\delta(y)-\delta(y-\frac{\pi R}{2}))
\,\epsilon^{\mu\nu\rho\sigma}{\rm str}(T^q(A_\nu F_{\rho\sigma}
+F_{\rho\sigma}A_\nu-A_\nu A_\rho A_\sigma)) \nonumber \\
U^{q={\hat a}}_\mu&=&-\frac{1}{96\pi^2}\delta(y)
\,\epsilon^{\mu\nu\rho\sigma}{\rm str}(T^q(A_\nu F_{\rho\sigma}
+F_{\rho\sigma}A_\nu-A_\nu A_\rho A_\sigma)).
\end{eqnarray}
For Type II, only the former one in the above is needed. 
Consequently, the CS term contributes to the anomaly for the 
5D covariant gauge current in Type I as 
\begin{eqnarray}
(D_M J^M)^{q_1=(a,i,B)}
&=&-\frac{1}{64\pi^2}(\delta(y)-\delta(y-\frac{\pi R}{2}))
\sum_{q_2,q_3=(b,j,B)}{\rm str}(T^{q_1}T^{q_2}T^{q_3})
F^{q_2}_{\mu\nu}F^{q_3\mu\nu}\nonumber \\
&-&\frac{1}{64\pi^2}\delta(y)
\sum_{q_2,q_3={\hat a}}{\rm str}(T^{q_1}T^{q_2}T^{q_3})
F^{q_2}_{\mu\nu}F^{q_3\mu\nu}, \label{csanomaly1}\\
(D_M J^M)^{q_1={\hat a}}&=&-\frac{1}{64\pi^2}\delta(y)
\sum_{q_2q_3={\hat b}(a,i,B)}{\rm str}(T^{q_1}T^{q_2}T^{q_3})
F^{q_2}_{\mu\nu}F^{q_3\mu\nu},\label{csanomaly2}
\end{eqnarray}
and in Type II as
\begin{eqnarray}
(D_M J^M)^{q_1=(a,i,B)}
&=&-\frac{1}{64\pi^2}(\delta(y)-\delta(y-\frac{\pi R}{2}))
\sum_{q_2,q_3=(b,j,B)}{\rm str}(T^{q_1}T^{q_2}T^{q_3})
F^{q_2}_{\mu\nu}F^{q_3\mu\nu} \nonumber \\ \\
(D_M J^M)^{q_1={\hat a}}&=&0
\end{eqnarray}
where $q_{1,2,3}$ run the bulk group indices.
It turns out that the CS contributions to the anomalies exactly cancel the 
localized gauge anomalies on the boundaries, 
eqs.~(\ref{bbanomaly1})-(\ref{bbanomaly4}) in Type I. On the other hand, 
in Type II, the CS term with a correct overall sign also exactly cancels 
the localized gauge anomalies, eqs.~(\ref{type2a}) and (\ref{type2b}), 
only in the case 
with two brane fields at different branes but not at the same brane.  
In the case with two brane fields at the same brane, we find that 
the bulk CS term is
not sufficient for cancelling all localized
anomalies from the bulk and brane matter fields.

\section{Fayet-Iliopoulos terms}

In Type I, the only place where the $U(1)$-graviton-graviton 
anomalies 
could appear is the fixed point $y=\pi R/2$ with
the local gauge group including a $U(1)$ gauge factor. On the other hand, 
in Type II, the mixed gravitational anomalies could appear on both fixed
points since their unbroken gauge group has a $U(1)$ gauge factor. As argued 
in the literature\cite{kkl}, however,
there is no gravitational counterpart $A\wedge R\wedge R$ 
of the 5D Chern-Simons term since the non-abelian gauge fields propagate 
in the bulk. 

It has been shown that the gravitational anomalies at $y=\pi R/2$
in Type I indeed cancel between the bulk and brane contributions 
without the need of a bulk Chern-Simons term\cite{kkl}. 
On the other hand, in Type II, it is only for two brane fields 
at different branes that there is no localized gravitational anomalies.
Then, since both gravitational anomalies
and FI terms are proportional to the common factor ${\rm Tr}(q)$, 
where $q$ is the $U(1)$ charge operator, it seems that
the absence of the gravitational anomalies should guarantee
the absence of the FI terms which could also exist at $y=\pi R/2$ in Type I
and at both fixed points in Type II. This is
the requirement for the stability of the 4D supersymmetric theory. 

In the orbifold models with an unbroken $U(1)$, however, it has been shown that 
the localized FI terms can be induced from a bulk field without breaking the 4D
supersymmetry\cite{scrucca,barbieri,nilles,pomarol}.
In this section, we present the explicit computation of
the Fayet-Iliopoulos(FI) terms\cite{nilles2,nilles}
for our set of bulk and brane fields in our model. 

The relevant part of the action for bulk($h,h^c$)
and brane($h_b$) scalar fields
for performing the FI term calculation in Type I is given by
\begin{eqnarray}
S&=&\int d^4 x\int^{2\pi R}_0 dy
\bigg[|\partial_M h|^2+|\partial_M h^{c\dagger}|^2
+g D^B(h^\dagger T^B h-h^c T^B h^{c\dagger}) \nonumber \\
&+&\delta(y-\frac{\pi R}{2})\bigg(|\partial_\mu h_b|^2
+gD^B h^\dagger_b q_b h_b\bigg)\bigg]
\end{eqnarray}
where $D^B$ imply the auxiliary field for the unbroken $U(1)$.
Denoting the bulk scalar fields as $h=(h_{++},h_{+-})^T$ and
$h^c=(h_{--},h_{-+})$, let us expand those in terms of bulk eigenmodes as
\begin{eqnarray}
h_{\pm\pm}(x,y)&=&\sum_n h_{(\pm\pm) n}(x)\xi^{(\pm\pm)}_n(y),\\
h_{\pm\mp}(x,y)&=&\sum_n h_{(\pm\mp) n}(x)\xi^{(\pm\mp)}_n(y).
\end{eqnarray}
As in the anomaly computation, inserting the above mode expansions 
in the 5D action gives
\begin{eqnarray}
S&=&\int d^4x\bigg[\sum_{\alpha,\beta=\pm}\sum_{m,n}
-h^\dagger_{(\alpha\beta) m}(x)
\bigg((\Box_4+M_n)\delta_{mn}
-gq_{\alpha\beta}D^{B(\alpha\beta)}_{mn}(x)\bigg) h_{(\alpha\beta) n}(x)
\nonumber \\
&-&h^\dagger_b(x)\bigg(\Box_4-gq_{--}
D^B(x,y=\frac{\pi R}{2})\bigg)h_b(x)\bigg]
\end{eqnarray}
where
\begin{eqnarray}
D^{B(\alpha\beta)}_{mn}(x)=\int^{2\pi R}_0 dy\,
\xi^{(\alpha\beta)}_m(y)\xi^{(\alpha\beta)}_n(y)D^B(x,y)
\end{eqnarray}
and
${\rm Tr}(T^B)=Kq_{++}+Nq_{+-}=0$, $q_{-+}=-q_{+-}$, $q_{--}=-q_{++}$,
and the introduction of a brane $\bar K$-plet with $q_b=q_{--}$ is understood.

From the one-loop tadpole diagram for the KK modes of auxiliary field $D^B$,
we can get the bulk and brane field contributions to the FI term
with the cutoff $\Lambda$ regularization as follows
\begin{eqnarray}
F(x)=F_{bulk}(x)+F_{brane}(x)
\end{eqnarray}
where
\begin{eqnarray}
F_{bulk}(x)=\sum_n \sum_{\alpha\beta}q_{\alpha\beta}T_n
D^{B(\alpha\beta)}_{nn}(x)
\end{eqnarray}
with
\begin{eqnarray}
T_n=ig\int\frac{d^4p}{(2\pi)^4}\frac{1}{p^2-M^2_n}
=\frac{g}{16\pi^2}\bigg(\Lambda^2-M^2_n\ln\frac{\Lambda^2+M^2_n}{M^2_n}\bigg),
\end{eqnarray}
and
\begin{eqnarray}
F_{brane}(x)=igKq_{--}D^B(x,y=\frac{\pi R}{2})\int\frac{d^4p}{(2\pi)^4}
\frac{1}{p^2}=\frac{gKq_{--}}{16\pi^2}\Lambda^2 D^B(x,y=\frac{\pi R}{2}).
\end{eqnarray}
Then, when we write the FI term in terms of the 5D field $D^B(x,y)$ as
\begin{eqnarray}
F(x)=\int^{2\pi R}_0dy\,f(y)D^B(x,y),
\end{eqnarray}
we make an inverse Fourier-transformation for the auxiliary field to
obtain the bulk profile for the FI term as
\begin{eqnarray}
f(y)=f_{even}(y)+f_{odd}(y)+f_{brane}(y)
\end{eqnarray}
where
\begin{eqnarray}
f_{even}(y)&=&gKq_{++}\sum_n T_n [|\xi^{(++)}_n|^2-|\xi^{(--)}_n|^2]\nonumber
\\
&=&\frac{gKq_{++}}{16\pi^2}\bigg[\frac{1}{2}\Lambda^2(\delta(y)
+\delta(y-\frac{\pi R}{2}))
+\frac{1}{4}\ln\frac{\Lambda}{\mu}(\delta^{\prime\prime}(y)
+\delta^{\prime\prime}(y-\frac{\pi R}{2}))\bigg],\\
f_{odd}(y)&=&gNq_{+-}\sum_n T_n [|\xi^{(+-)}_n|^2-|\xi^{(-+)}_n|^2]\nonumber
\\
&=&\frac{gNq_{+-}}{16\pi^2}\bigg[\frac{1}{2}\Lambda^2(\delta(y)
-\delta(y-\frac{\pi R}{2}))
+\frac{1}{4}\ln\frac{\Lambda}{\mu}(\delta^{\prime\prime}(y)
-\delta^{\prime\prime}(y-\frac{\pi R}{2}))\bigg], \\
f_{brane}(y)&=&\frac{gKq_{--}}{16\pi^2}\Lambda^2\delta(y-\frac{\pi R}{2}).
\end{eqnarray}
Here, prime denotes the derivative
with respect to the extra dimension coordinate. Consequently, the resultant
FI term in Type I is given by
\begin{eqnarray}
f(y)&=&\frac{g{\rm Tr}(T^B)}{32\pi^2}\bigg[\Lambda^2(\delta(y)
+\delta(y-\frac{\pi R}{2}))+\frac{1}{2}\ln\frac{\Lambda}{\mu}
\delta^{\prime\prime}(y)\bigg]+\frac{gKq_{++}}{32\pi^2}
\ln\frac{\Lambda}{\mu}\delta^{\prime\prime}(y-\frac{\pi R}{2}) \nonumber \\
&=&\frac{gKq_{++}}{32\pi^2}
\ln\frac{\Lambda}{\mu}\delta^{\prime\prime}(y-\frac{\pi R}{2})
\end{eqnarray}
where we used ${\rm Tr}(T^B)=0$ in the last line. 
We note that there is no FI term at $y=0$ with the full bulk gauge group,
which is as expected because there
is no $U(1)$ factor at this fixed point. Moreover, we find that there is no
conventional FI term with quadratic divergence even at $y=\pi R/2$
with a $U(1)$ factor, which is
consistent with the absence of mixed gravitational anomalies as argued
in $\cite{kkl}$. However, there exists a non-vanishing FI term
with logarithmic divergence at $y=\pi R/2$.

Likewise, following the same procedure, we find that the localized FI terms 
in Type II 
depend on the location of brane fields: with both $\bar K$-plet and $N$-plet 
at the same brane $y=0$($y=\pi R/2$) as
\begin{eqnarray}
f(y)=\pm \frac{gKq_{++}}{16\pi^2}\Lambda^2
\bigg(\delta(y)-\delta(y-\frac{\pi R}{2})\bigg)
+\frac{gKq_{++}}{32\pi^2}\ln\frac{\Lambda}{\mu}
\bigg(\delta^{\prime\prime}(y)+\delta^{\prime\prime}(y-\frac{\pi R}{2})\bigg),
\end{eqnarray}
and with a $\bar K$-plet at $y=0$($y=\pi R/2$) and a $N$-plet 
at $y=\pi R/2$($y=0$) as
\begin{eqnarray}
f(y)=\frac{gKq_{++}}{32\pi^2}\ln\frac{\Lambda}{\mu}
\bigg(\delta^{\prime\prime}(y)+\delta^{\prime\prime}(y-\frac{\pi R}{2})\bigg).
\end{eqnarray}
In the case with two brane fields at the same brane, there appear 
quadratic FI terms as well as log FI terms. However, in this case, 
even if the integrated FI terms are zero, the localized FI terms with quadratic
divergence would not give rise to a consistent theory at low energies 
since there is no gravitational CS term to cancel the mixed
gravitational anomalies, $\pm Kq_{++}$ at $y=0$ and $\mp Kq_{++}$ at $y=\pi R$.
On the other hand, in the case with two brane fields at different branes,
there are only log divergent FI terms at both fixed points which maintain 
the 4D supersymmetry and the mass spectrum of a bulk multiplet\cite{nilles}.

\section{Conclusion}
We considered the breaking of the 5D non-abelian gauge symmetry 
on $S^1/(Z_2\times Z'_2)$ orbifold. 
Then, we presented the localized gauge anomalies coming from a bulk 
fundamental field through the explicit KK mode decomposition of the 5D fields. 
In the orbifold with gauge symmetry breaking, there are fixed points with their
own local gauge symmetries. Thus, there is the possibility of embedding
some incomplete multiplets at the fixed point with unbroken gauge group,
which can be sometimes phenomenologically preferred. The incomplete brane 
multiplet we considered can be realized from a bulk muliplet in the field 
theoretic limit. Therefore, we have shown that
the 4D anomaly combination of a brane field and a bulk zero mode  
does not have the localized gauge anomalies 
up to the addition of a Chern-Simon 5-form with some jumping coefficient, 
which could be regarded as the effects of the bulk heavy modes as in the 
abelian gauge theory on $S^1/Z_2$. However, for the brane fields
assigned at the same fixed points in models of Type II, we found that 
it is not possible to cancel the localized
anomalies only with the CS term even if there is no 4D anomaly.. 
Then, we also found a nonzero log FI term at the fixed point with $H$ 
in both types of orbifold boundary conditions we considered.

\bigskip

\acknowledgments
I would like to thank H. D. Kim, J. E. Kim, H. P. Nilles,
M. Olechowski and M. Walter for reading the draft and useful discussions.
This work is supported in part by the KOSEF Sundo Grant, and by the
European Community's Human Potential Programme under contracts
HPRN-CT-2000-00131 Quantum Spacetime, HPRN-CT-2000-00148 Physics Across the
Present Energy Frontier and HPRN-CT-2000-00152 Supersymmetry and the Early
Universe. I was supported by priority grant 1096 of the Deutsche
Forschungsgemeinschaft.


\begin{thebibliography}{999}

\bibitem{kawamura} Y. Kawamura, {\it Triplet doublet splitting, prton decay 
and extra dimension}, \ptp{105}{2001}{999} [hep-ph/0012125];
G. Altarelli and F. Feruglio, {\it $SU(5)$ grand unification in extra 
dimensions and proton decay}, \plb{511}{2001}{257} [hep-ph/0102301].

\bibitem{pointgroup}
L. J. Hall and Y. Nomura, {\it Gauge unification in higher dimensions}, 
\prd{64}{2001}{055003} [hep-ph/0103125];
A. Hebecker and J. March-Russell, {\it A minimal $S^1/(Z_2\times Z'_2)$ 
orbifold GUT}, \npb{613}{2001}{3} [hep-ph/0106166];
A. Hebecker and J. March-Russell, {\it The structure of GUT breaking
by orbifolding}, \npb{625}{2002}{128} [hep-ph/0107039];
L. J. Hall and Y. Nomura, {\it Gauge coupling unification from unified 
theories in higher dimensions}, \prd{65}{2002}{125012} [hep-ph/0111068];
L. J. Hall and Y. Nomura, {\it A complete theory of grand unification 
in five-dimensions}, \prd{66}{2002}{075004} [hep-ph/0205067].

\bibitem{kkl0}
H. D. Kim, J. E. Kim and H. M. Lee, {\it Top-bottom mass hierarchy, s-$\mu$ 
puzzle and gauge coupling unification with split multiplets}, 
\epjc{24}{2002}{159} [hep-ph/0112094].

\bibitem{su3}
S. Dimopoulos and D. E. Kaplan, {\it The weak mixing angle from an $SU(3)$
symmetry at a TeV}, \plb{531}{2002}{127} [hep-ph/0201148];
S. Dimopoulos, D. E. Kaplan and N. Weiner,
{\it Electroweak unification into a five-dimensional $SU(3)$ at a TeV}, 
\plb{534}{2002}{124} [hep-ph/0202136];
L. J. Hall and Y. Nomura, {\it Unification of weak and hypercharge interactions
at the TeV scale}, \plb{532}{2002}{111} [hep-ph/0202107];
T. -J Li and W. Liao, {\it Weak mixing angle and the $SU(3)_c\times SU(3)$ 
model on $M^4\times S^1/(Z_2\times Z'_2)$}, 
\plb{545}{2002}{147} [hep-ph/0202090].

\bibitem{kkl}
H. D. Kim, J. E. Kim and H. M. Lee, {\it TeV scale 5D $SU(3)_W$ unification
and the fixed point anomaly cancellation with chiral split multiplets}, 
\jhep{06}{2002}{048} [hep-th/0204132]; 
H. M. Lee, {\it 5D $SU(3)_W$ unification at TeV and cancellation of localized
gauge anomalies with split multiplets}, hep-th/0210177.

\bibitem{anton} I. Antoniadis, {\it A possible new dimension at a few TeV}, 
\plb{246}{1990}{377};
I. Antoniadis and K. Benakli, {\it Limits on extra dimensions in orbifold
compactifications of superstrings}, \plb{326}{1994}{69}.

\bibitem{scrucca}
C. A. Scrucca, M. Serone, L. Silvestrini and F. Zwirner,
{\it Anomalies in orifold field theories}, 
\plb{525}{2002}{169} [hep-th/0110073].

\bibitem{pilo}
L. Pilo and A. Riotto, {\it On anomalies in orbifold theories}, 
\plb{546}{2002}{135} [hep-th/0202144].

\bibitem{barbieri}
R. Barbieri, R. Contino, P. Creminelli, R. Rattazzi and C. A. Scrucca,
{\it Anomalies, Fayet-Iliopoulos terms and the consistency of orbifold field
theories}, \prd{66}{2002}{024025} [hep-th/0203039].

\bibitem{scrucca2}
C. A. Scrucca, M. Serone and T. Trapletti, {\it Open string models with 
Scherk-Schwarz SUSY breaking and localized anomalies}, 
\npb{635}{2002}{33} [hep-th/0203190].

\bibitem{nilles}
S. G. Nibbelink, H. P. Nilles and M. Olechowski, {\it Instabilities of bulk
fields and anomalies on orbifolds}, \npb{640}{2002}{171} [hep-th/0205012].

\bibitem{ah} N. Arkani-Hamed, A. G. Cohen and H. Georgi,
{\it Anomalies on orbifolds}, \plb{516}{2001}{395} [hep-th/0103135].

\bibitem{ch}
C. G. Callan and J. A. Harvey, {\it Anomalies and fermion zero modes
on strings and domain walls}, \npb{250}{1985}{427}.

\bibitem{nilles2} D. M. Ghilencea, S. Groot Nibbelink and H. P. Nilles,
{\it Gauge corrections and FI term in 5D KK theories}, 
\npb{619}{2001}{385} [hep-th/0108184].

\bibitem{peskin}
E. A. Mirabelli and M. E. Peskin, {\it Transmission of supersymmetry breaking
from a four-dimensional boundary}, \prd{58}{1998}{065002} [hep-th/9712214].

\bibitem{ah2}
N. Arkani-Hamed, T. Gregoire and J. Walker, {\it Higher dimensional 
supersymmetry in 4D superspace}, \jhep{0203}{2002}{055} [hep-th/0101233].

\bibitem{pomarol}
D. Marti and A. Pomarol, {\it Fayet-Iliopoulos terms in 5D theories and
their phenomenological implications}, \prd{66}{2002}{125005} [hep-ph/0205034].

\bibitem{hebecker}
A. Hebecker and J. March-Russell, {\it The flavor hierarchy and seesaw 
neutrinos from bulk masses in 5D orbifold GUTs}, 
\plb{541}{2002}{338} [hep-ph/0205143].

\bibitem{bardeen} W. Bardeen and B. Zumino, {\it Consistent and covariant
anomalies in gauge and gravitational theories}, \npb{244}{1984}{421}.





\end{thebibliography}
\end{document}